\documentstyle[preprint,aps,prl,epsf]{revtex}
\begin{document}


 \begin{flushright}
 SLAC--PUB--8951   \\
 \vspace*{-3mm}
 SCIPP--01/38      \\
 \vspace*{-3mm}
 November 2001
 \end{flushright}

\begin{center}
\begin{large}
    {\bf Improved Direct Measurement of $A_{b}$ and $A_c$
       at the $Z^0$ Pole \\}
\vspace*{-4mm}
    {\bf Using a Lepton Tag$^*$}

\vspace*{8mm}
{\bf The SLD Collaboration$^{**}$}
\end{large}

\vspace*{2mm}

Stanford Linear Accelerator Center    \\
\vspace*{-2mm}
Stanford University, Stanford, CA 94309

\vspace*{12mm}
{\bf Abstract}
\end{center}

The parity violation parameters $A_b$ and $A_c$ of the $Zb{\overline b}$
and $Zc\bar{c}$
couplings have been measured directly, using the polar angle
dependence of the polarized cross sections at the $Z^0$ pole.
Bottom and charmed hadrons
were tagged via their semileptonic decays.
Both the electron and muon analyses take advantage of new
multivariate techniques to increase the analyzing power.
Based on the 1993-98 SLD sample of 550,000 $Z^0$ decays produced with highly
polarized electron beams we measure
$ A_b = 0.919 \pm 0.030_{stat} \pm 0.024_{syst} $, and
$ A_c = 0.583 \pm 0.055_{stat} \pm 0.055_{syst} $.

\vspace*{28mm}
\begin{center}
{\it Submitted to Physical Review Letters}
\end{center}

\vspace*{12mm}

$^*$ Work supported in part by Department of Energy contract
 DE-AC03-76SF00515

\vfill
\eject

\narrowtext
Parity violation in the $Zf\bar{f}$ coupling can be measured via the
observables  
$A_f = 2v_fa_f/(v^2_f+a^2_f),$ 
where $v_f$ and $a_f$ represent the vector and
axial vector couplings to fermion $f$.  
In particular, for $f = b$, $A_b$ is largely independent of propagator
effects that modify the effective weak mixing angle, and thus provides
an unambiguous test of the Standard Model.

The Born-level differential cross section for the process 
$e^{+} e^{-} \rightarrow  Z^0 \rightarrow f\bar{f}$ is
\begin{equation}
{d \sigma_f} \, / \, {dz} \propto
(1-A_e P_e) (1+z^2) + 2A_f (A_e - P_e) z \, ,
\end{equation}
where $P_e$ is the $e^-$ beam longitudinal polarization
($P_e > 0$ for right-handed (R) polarization)
and $z$ is the cosine of the polar angle of the outgoing fermion
with respect to the incident electron.
The ability to modulate the sign of $P_e$
allows the final-state quark coupling
$A_f$ to be extracted independently of
$A_e$ from a fit to the differential cross section.
Thus, the measurements of $A_f$ described here are
unique, and complementary to other electroweak measurements performed
at the $Z^0$ pole\cite{EWWG}.

This Letter reports the results of the 1996-98 SLD lepton tag analysis,
for which identified electrons and muons were used to tag the flavor
of the underlying heavy quark. The data sample used in this analysis
is roughly three times larger than
that of previously reported results~\cite{lepton99}.
Further statistical and systematic advantage is provided by
improvements to the data analysis
which take advantage of the precise information provided
by the new vertex detector (VXD3)~\cite{vxd3} that was installed just
prior to the 1996 data run.

The Stanford Linear Collider (SLC) and its operation with a polarized
electron beam have been described elsewhere \cite{SLC}.
During the 1996-98 run, the
SLC Large Detector (SLD)~\cite{SLD} recorded an integrated
luminosity of $14.0$ pb$^{-1}$
at a mean center of mass energy of 91.24 GeV,
with a luminosity-weighted electron beam polarization of
$|P_e| = 0.7336 \pm 0.0038 $~\cite{ALR}.

Charged particle tracks are reconstructed in the Central 
Drift Chamber (CDC)
and the CCD-based vertex detector in a uniform axial
magnetic field of 0.6T. 
For the 1996-98 data, the combined CDC and VXD3
impact parameter resolution in the
transverse (longitudinal) direction with respect to the beam is
7.7 (9.6) $\mu$m at high momentum, and 34 (34) $\mu$m at
$p_{\perp}\sqrt{\sin \theta}$ = 1 GeV/c$^2$, where
$p_{\perp}$ and $\theta$ are the
momentum transverse to and angle relative to the electron beam
direction.
The Liquid Argon Calorimeter (LAC) measures the energy and shower
profile of charged and neutral particles with an electromagnetic
 energy resolution of
$ {{\sigma_E} / {E}} = {{15\%} / {\sqrt{E(GeV)}}} $ and is used in the
electron identification.  The Warm Iron Calorimeter (WIC) detects charged
particles that penetrate the 3.5 interaction lengths of the LAC and 
magnet coil. The Cherenkov Ring Imaging Detector (CRID) measures the
velocity of charged tracks in the region $|\cos\theta| < 0.68$
using the number and angle of Cherenkov photons emitted
in liquid and gaseous radiators;
electrons are well separated from pions in the region between 2 and 5 GeV/c, while
pion (kaon) rejection reduces backgrounds to the muon sample
in the region $2 < p < 5$ ($2 < p < 15$) GeV/c.

The axis of the jet nearest in angle to the lepton candidate is used to
approximate $z$, the cosine of the polar angle of the underlying quark.
Jets are formed from calorimeter energy clusters
(including any associated with the lepton candidate)
using the JADE algorithm \cite{JADE}
with parameter $y_{cut} = 0.005$.
The analyses presented here make substantial
use of `secondary' decay vertices which are displaced from the
primary interaction point, identified via the ZVTOP topological
vertexing algorithm~\cite{zvtop}, as well as the
invariant mass of the tracks comprising the secondary
vertex (`vertex mass'), corrected to account for unmeasured neutral
particles~\cite{ptcor}.


The selection of electron and
muon candidates with $p > 2$ GeV/c in hadronic $Z^0$ decays has been described
previously~\cite{lepton99}.
Electrons are identified with both LAC and CRID information
for CDC tracks in the angular range
$|\cos\theta| < 0.72$. Electrons from photon conversions are
recognized and removed with 73\% efficiency.
WIC information is also included for muons,
providing an essential measurement of their penetration.
Muons are identified in the angular region $|\cos\theta| < 0.70$,
although the identification efficiency falls rapidly for
$|\cos\theta| > 0.60$ due to the limited angular coverage of the WIC.
To reduce backgrounds from misidentification,
the 29\% of
events containing electron candidates that had no reconstructed secondary
vertices were removed from the sample, precluding the use of the electron
sample for the measurement of $A_c$.

For $p > 2$ GeV/c,
Monte Carlo (MC) studies indicate efficiencies (purities) of 64\% (64\%)
and 81\% (68\%) for the electron and muon samples, respectively, where
the remaining
electrons from photon conversion account for 5\% of the 12862 electron candidates.
In the case of the muon sample (21199 candidates), the background is due both to
misidentification (8\% of muon candidates) and to real muons from light
hadron decays (25\%). In both cases, the MC simulation has been
verified with a control sample of pions from $K^0_S \rightarrow \pi^+\pi^-$
decays.  The fraction of such pions misidentified as electrons is
($1.02 \pm 0.06$)\%, consistent with the MC expectation of ($1.06 \pm 0.03$)\%.
For muons, the measured pion misidentification fraction is ($0.342 \pm 0.028$)\%,
somewhat higher than the MC expectation of ($0.279 \pm 0.012$)\%.
This difference has been accounted for by raising the background level
in the maximum likelihood fit to the muon sample by $(20 \pm 10)\%$ of itself.

The sample of events containing
identified leptons is composed of the following event types
(charge conjugates implied):
\hbox{$Z^0 \rightarrow b\bar{b}, b \rightarrow l$ (`$bl$');}
\hbox{$Z^0 \rightarrow b\bar{b}, b \rightarrow \bar{c}
                            \rightarrow l$ (`$b\bar{c}l$');}
\hbox{$Z^0 \rightarrow b\bar{b}, \bar{b} \rightarrow \bar{c}
                            \rightarrow l$ (`$\bar{b}\bar{c}l$');}
\hbox{$Z^0 \rightarrow c\bar{c}, \bar{c} \rightarrow l$ (`$\bar{c}l$');}
and background from light hadron and vector meson
decays, photon conversions,
and misidentified hadrons (`$bk$').  

Identification of electron candidate event types is based on
the values of eight discriminating
variables~\cite{Jorge}:
track momentum ($p$), momentum transverse to the nearest
jet ($p_t$), the estimate of the underlying B hadron boost\cite{frag}
and when
available, same hemisphere secondary vertex mass,
opposite hemisphere vertex mass,
same hemisphere vertex momentum resultant, same hemisphere
vertex significance (separation $D$ between the
interaction point and secondary vertex, divided by its uncertainty), and $L/D$
(where $L$ is the distance from the interaction point to the point on
the secondary vertex trajectory closest to the electron candidate
trajectory).
These variables are used as inputs to an Artificial Neural Network
with three output nodes $N_{bl}$, $N_{bcl}$, and $N_{cl}$, optimized for the
$bl$, $b\bar{c}l+\bar{b}\bar{c}l$, and $\bar{c}l$ signals, respectively.
Event type probabilities are
estimated according to the composition of MC electron candidate
events with similar output node values.
The measured and simulated distributions of the three output node variables
are compared in Figure~\ref{nnout}.

The Neural Network is trained on the SLD MC sample of hadronic
$Z^0$ decays,
generated with JETSET 7.4\cite{LUND}.
Semileptonic decays of $B$ mesons are generated according to the
ISGW formalism \cite{ISGW} with a 23\% $D^{**}$ fraction,
while semileptonic decays of $D$ mesons are simulated according to 
branching ratios reported by the Particle Data Group~\cite{pdb}.
Experimental constraints are provided by
the $B \rightarrow l$ and $B \rightarrow D$ inclusive momentum spectra
measured by the CLEO collaboration \cite{CLEO,ISGW2} and
the $D \rightarrow l$ momentum spectrum measured by
the DELCO collaboration \cite{DELCO}.
The detailed simulation of the SLD detector response has been realized using
GEANT \cite{GEANT}.

\begin{figure}
 \bigskip
 \epsfxsize=3.3in
 \epsfysize=3.0in
 \centerline{
   \epsfbox{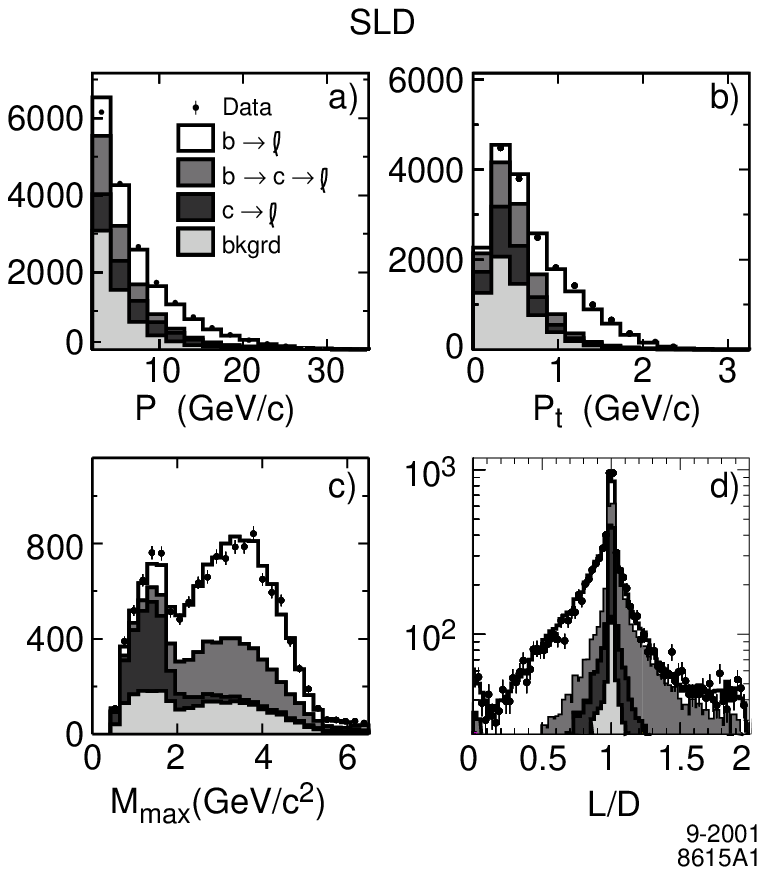}}
 \caption{Distributions of data and MC muon candidates.}
 \label{muondist}
 \end{figure}

Muon candidate event type probabilities are
estimated according to the composition
of MC muon candidate events with similar values of the following
discriminating variables~\cite{Giulia}:
$p$, $p_t$, and, when available, $L/D$
and $M_{max}$, the largest of the secondary vertex invariant masses. 
The measured and simulated distributions of these variables
are compared in Figure~\ref{muondist}.


A maximum likelihood analysis of all selected hadronic $Z^0$ events containing
lepton candidates is used to determine $A_{b}$ and $A_c$.
The likelihood function contains the following probability term
for each lepton, with measured charge sign $Q$:
\widetext
\begin{eqnarray}
P\,(P_e,z; \, A_b) & \propto &
  \left\{ (1+z^2)(1-A_eP_e)
  -2Q(A_e-P_e)\left[(f_{bl}(1-2\bar \chi_b)-f_{\bar{b}\bar{c}l}(1-2\bar
     \chi_{\bar{b}\bar{c}}) \right. \right.  \nonumber \\
& & \left. \left. \mbox{} + f_{b\bar c l}(1-2\bar\chi_{b\bar c}))
 (1-\Delta^b_{QCD}(z))A_b
 - f_{\bar{c}l}(1-\Delta^c_{QCD}(z))A_c+f_{bk}A_{bk}\right]z\right\}.
\end{eqnarray}
\narrowtext
The lepton source fractions $f_{bl}$, $f_{\bar{b}\bar{c}l}$, $f_{b\bar{c}l}$,
$f_{\bar{c}l}$,
and $f_{bk}$
are functions of the three neural net output node values (electron candidates)
or the four discriminating variables (muon candidates).
For the fit to muon candidates,
both $A_b$ and $A_c$ are left as free parameters, whereas $A_c$ is
fixed to its SM value (see Table I) for the fit to electron candidates.

Correction factors $(1-2 \bar \chi_x)$, where $\chi_x$ is the mixed fraction
for lepton source $x$,
are applied to  b-quark lepton
sources to account for asymmetry dilution due to $B^0\bar{B}^0$ mixing.
The value of
$\bar \chi_b$ is taken from LEP measurements of the average
mixing in semileptonic $B$ decays\cite{EWWG},
but must be corrected to take into account selection and fitting bias,
including that due to
the enhanced likelihood for $bcl$ cascade leptons to have come from
a $B$ meson which has mixed~\cite{bclmix}. For the electron sample,
MC studies indicated that the mixing probability $\bar \chi_b$ 
for $bl$ decays was independent of
of the value of the NN output parameters, but was increased by a relative
1.7\% overall by the bias of the vertex requirement towards the selection
of $B^0$ over $B^{\pm}$ decays. For the muon sample, the effective values
of $\bar\chi_{\bar{b}\bar{c}}$ and $\bar\chi_{b\bar c}$ were evaluated
on an event-by-event basis, based on MC events with values of the
muon-sample discriminating variables close to those of the given data event.


The asymmetry in the background $A_{bk}$ is parameterized as a function of
$p$ and $p_t$. For the electron sample, the parameterization
is determined from tracks in the data not identified as
leptons. For the muon sample, MC studies indicated a substantial difference
between the true background asymmetry and that of non-leptonic tracks, and
so the background asymmetry parameterization was determined directly from the
MC simulation.

A $z$-dependent correction factor $(1-\Delta^f_{QCD}(z))$
is included in the likelihood function
to incorporate the effects of gluon radiation. Calculation of
the quantity $\Delta^f_{QCD}(z)$ has been performed by several
groups~\cite{QCDall}.
For an unbiased sample of $b\bar
b$ or $c\bar c$ events with $|z|<0.7$, 
correcting for this effect increases the measured asymmetry by $\sim 3\%$
overall. However, a MC simulation of the analysis chain indicates that
biases which favor $q\bar{q}$ events over $q\bar{q}g$ events
mitigate the effects of leading order gluon
radiation by about 30\%.
Effects due to gluon splitting to $b \bar b$ and
$c \bar c$ have been estimated by rescaling the JETSET simulation to
world average gluon splitting measurements~\cite{split}. 
Additional radiative effects, such as those due to initial-state radiation
and $\gamma/Z$
interference, lead to a further correction of $-0.2\%$ ($-0.1\%$) on
the value of $A_b$ ($A_c$).


A list of systematic errors is shown in Table~\ref{systable}. 
The purity of the separation of
$Z^0 \rightarrow b{\bar b}$ and $Z^0 \rightarrow c{\bar c}$ events
via secondary vertex information introduces an uncertainty
dominated by the efficiency of charged track
reconstruction, which has been constrained by reweighting
MC tracks by the ratio of the number of tracks in data and MC
as a function of $p$ and $p_t$. The ability of the L/D
variable to discriminate between $bl$ and $b{\bar c}l$ decays
is sensitive to the fraction of $B \rightarrow D {\bar D}$ decays,
which has been constrained from SLD data\cite{Giulia}.

For the 1996-98 muon sample, we find that 
$A_b = 0.938 \pm 0.044({\rm stat.}) \pm 0.024({\rm syst.})$ and
$A_c = 0.560 \pm 0.063({\rm stat.}) \pm 0.064({\rm syst.})$, with a statistical
correlation coefficient of 0.108.
For the corresponding electron sample, we find
$A_b = 0.896 \pm 0.050({\rm stat.}) \pm 0.028({\rm syst.}).$
Combined with the result of~\cite{lepton99}, we find
overall SLD average
results via semileptonic $B$ and $D$ hadron decay of
$$ A_b = 0.919 \pm 0.030({\rm stat.}) \pm 0.024({\rm syst.}) $$
$$ A_c = 0.583 \pm 0.055({\rm stat.}) \pm 0.055({\rm syst.}). $$


In conclusion, we have directly
measured the extent of parity violation in the
coupling of $Z^0$ bosons to $b$ and $c$ quarks using identified charged
leptons from semileptonic decays. The results presented here take
advantage of an additional sample of 400000 $Z^0$ decays, and
employ a new method of signal source separation, resulting in
substantial increases in
precision relative to previous measurements\cite{lepton99}. These results
are in agreement with
the Standard Model predictions $A_b = 0.935$ and
$A_c = 0.667$, which are insensitive to uncertainties in Standard Model
parameters such as the strong and electromagnetic coupling strengths, and the
top quark and Higgs boson masses.
           
We thank the staff of the SLAC accelerator department for their
outstanding efforts on our behalf. This work was supported by the
U.S. Department of Energy and National Science Foundation, the UK Particle
Physics and Astronomy Research Council, the Istituto Nazionale di Fisica
Nucleare of Italy and the Japan-US Cooperative Research Project on High
Energy Physics.

%

%
%

%
%

%
%

%
%
\widetext
\begin{table*}
\small
\begin{tabular}{llccc}
Source & Parameter variation & $\delta A_b (\mu) $ & $\delta A_b (e) $ & $\delta A_c (\mu) $ \\ \hline
Monte Carlo statistics  & Includes Neural Net training for $e$
                     &  $\pm$.005 & $\pm$.014  & $\pm$.023 \\
Jet axis simulation     & 10 mrad smearing
                     & $\pm$.002 & $\pm$.006  & $\pm$.002 \\
Background level        & $\pm 10\%$ relative
                     & $\pm$.003 & $\pm$.004  & $\pm$.010 \\
Background asymmetry    & $\pm 40\%$ relative
                     & $\mp$.002 & $\mp$.003  & $\pm$.007 \\
BR($Z^0 \rightarrow b\bar{b}$)     & $R_b = .2164 \pm .0007$
                     & $\mp$.000 & $\pm$.000    & $\pm$.001 \\
BR($Z^0 \rightarrow c\bar{c}$)     & $R_c = .1674 \pm .0038$
                     & $\pm$.001 & $\pm$.000  & $\mp$.008 \\
BR($b \rightarrow l$)     & $(10.62 \pm 0.17)\% $
                     & $\mp$.003 & $\mp$.003  & $\pm$.003 \\
BR($\bar{b} \rightarrow \bar{c} \rightarrow l$)     & $(8.07 \pm 0.25)\% $
                     & $\pm$.003 & $\pm$.003 & $\mp$.003 \\
BR($b \rightarrow \bar{c} \rightarrow l$)     & $(1.62 \pm 0.40)\% $
                     & $\mp$.006 & $\mp$.001 & $\pm$.011 \\
BR($b \rightarrow \tau \rightarrow l$)     & $(0.452 \pm 0.074)\% $
                     & $\mp$.003   & $\mp$.001   & $\mp$.002 \\
BR($b \rightarrow J / \psi   \rightarrow l$)     & $(0.07 \pm 0.02)\% $
                     & $\pm$.003 & $\pm$.002 & $\pm$.000 \\
BR($\bar{c} \rightarrow l$)     & $(9.85 \pm 0.32)\% $
                     & $\pm$.001 & $\pm$.001 & $\mp$.012 \\
$B$ lept. spect. - $D^{**}$ fr.
                     & $(23 \pm 10)\%$, $B^{+}$,$B^{0}$;
                       $(32 \pm 10)\%$, $B_{s}$
                     & $\pm$.003 & $\pm$.002 & $\pm$.001 \\
$D$ lept. spect.
                     & $ACCMM1 \: (^{+ACCMM2}_{-ACCMM3})$ \cite{ACCMM}\
                     & $\pm$.004 & $\pm$.004 & $\pm$.002 \\
$B_{s}$ fraction in $b\bar{b}$ event     
                     & $.115 \pm .050$
                     & $\pm$.001 & $\pm$.004 & $\mp$.001 \\
$\Lambda_{b}$ fraction in $b\bar{b}$ event     
                     & $.072 \pm .030$
                     & $\pm$.002 & $\pm$.002 & $\mp$.001 \\
$b$  fragmentation   & $\epsilon_b=.0045$-$.0075$\cite{LUND}
                     & $\pm$.001 & $\pm$.004 & $\pm$.002 \\
$c$  fragmentation   & $\epsilon_c=.045$-$.070$\cite{LUND}
                     & $\mp$.003 & $\mp$.000 & $\pm$.012 \\
Polarization         & $<\!P_e\!>= {73.4\pm 0.4}$
                     & $\mp$.005 & $\mp$.005 & $\mp$.003 \\
QCD Corrections      & $\alpha_S$, gluon splitting, selection bias
                     & $\pm$.005 & $\pm$.005 &$\pm$.005   \\
Gluon Splitting      &  $g_{cc} = (2.33 \pm 0.50)\%$;
                        $g_{bb} = (0.27 \pm 0.07)\%$
                     & $\pm$.001 &$\pm$.001  &$\pm$.002   \\
$B$ mixing $\chi_b$  & $\chi = .1186 \pm .0043 $
                     & $\pm$.010 & $\pm$.011 & $\pm$.000 \\
$N_{D^0}/N_{D+}$ in B Decay & $\pm 10\%$
                      & $\pm$.002  & $\pm$.001 &$\pm$.003  \\
B tag purity         & Track efficiency
                     & $\pm$.012 &$\pm$.014  &$\pm$.053  \\
L/D variable         &   Data/MC comparison
                     &$\pm$.002 &$\pm$.000  &$\pm$.005  \\
Neural Net Training  &
                     & ---        & $\pm .013$  &  ---       \\
$B \rightarrow D \bar{D} \rightarrow l$  &  $(11.5 \pm 2.5)\%$
                     &$\pm$.010   & $\pm .008$ & $\pm .003$  \\
$A_c$                & $0.667 \pm 0.030$
                     &  ---   &  $\pm$0.002  & --- \\
\hline
Total Systematic   & & $\pm$.024  & $\pm$.028  & $\pm$.064  \\
\end{tabular}
\caption{Systematic errors}
\label{systable}
\end{table*}

%
%
\narrowtext

\listoffigures
\listoftables

\vfill \eject

\begin{large}
\begin{center}
$^{**}$ {\bf The SLD Collaboration}
\end{center}
\end{large}

%
%
%
\begin{center}
\def\iAOMORI{$^{(1)}$}
\def\iBRI{$^{(2)}$}
\def\iBRUN{$^{(3)}$}
\def\iBU{$^{(4)}$}
\def\iCOLO{$^{(5)}$}
\def\iCSU{$^{(6)}$}
\def\iFERR{$^{(7)}$}
\def\iFRAS{$^{(8)}$}
\def\iJHU{$^{(9)}$}
\def\iLBL{$^{(10)}$}
\def\iMASS{$^{(11)}$}
\def\iMISSI{$^{(12)}$}
\def\iMIT{$^{(13)}$}
\def\iMOSCOW{$^{(14)}$}
\def\iNAGO{$^{(15)}$}
\def\iOREG{$^{(16)}$}
\def\iOXF{$^{(17)}$}
\def\iPERU{$^{(18)}$}
\def\iRAL{$^{(19)}$}
\def\iRUTG{$^{(20)}$}
\def\iSLAC{$^{(21)}$}
\def\iSOONG{$^{(22)}$}
\def\iTENN{$^{(23)}$}
\def\iTOHO{$^{(24)}$}
\def\iUCSB{$^{(25)}$}
\def\iUCSC{$^{(26)}$}
\def\iVAND{$^{(27)}$}
\def\iWASH{$^{(28)}$}
\def\iWISC{$^{(29)}$}
\def\iYALE{$^{(30)}$}

  \baselineskip=.75\baselineskip
\mbox{Kenji Abe\unskip,\iNAGO}
\mbox{Koya Abe\unskip,\iTOHO}
\mbox{T. Abe\unskip,\iSLAC}
\mbox{I. Adam\unskip,\iSLAC}
\mbox{H. Akimoto\unskip,\iSLAC}
\mbox{D. Aston\unskip,\iSLAC}
\mbox{K.G. Baird\unskip,\iMASS}
\mbox{C. Baltay\unskip,\iYALE}
\mbox{H.R. Band\unskip,\iWISC}
\mbox{T.L. Barklow\unskip,\iSLAC}
\mbox{J.M. Bauer\unskip,\iMISSI}
\mbox{G. Bellodi\unskip,\iOXF}
\mbox{R. Berger\unskip,\iSLAC}
\mbox{G. Blaylock\unskip,\iMASS}
\mbox{J.R. Bogart\unskip,\iSLAC}
\mbox{G.R. Bower\unskip,\iSLAC}
\mbox{J.E. Brau\unskip,\iOREG}
\mbox{M. Breidenbach\unskip,\iSLAC}
\mbox{W.M. Bugg\unskip,\iTENN}
\mbox{D. Burke\unskip,\iSLAC}
\mbox{T.H. Burnett\unskip,\iWASH}
\mbox{P.N. Burrows\unskip,\iOXF}
\mbox{A. Calcaterra\unskip,\iFRAS}
\mbox{R. Cassell\unskip,\iSLAC}
\mbox{A. Chou\unskip,\iSLAC}
\mbox{H.O. Cohn\unskip,\iTENN}
\mbox{J.A. Coller\unskip,\iBU}
\mbox{M.R. Convery\unskip,\iSLAC}
\mbox{V. Cook\unskip,\iWASH}
\mbox{R.F. Cowan\unskip,\iMIT}
\mbox{G. Crawford\unskip,\iSLAC}
\mbox{C.J.S. Damerell\unskip,\iRAL}
\mbox{M. Daoudi\unskip,\iSLAC}
\mbox{N. de Groot\unskip,\iBRI}
\mbox{R. de Sangro\unskip,\iFRAS}
\mbox{D.N. Dong\unskip,\iSLAC}
\mbox{M. Doser\unskip,\iSLAC}
\mbox{R. Dubois\unskip,\iSLAC}
\mbox{I. Erofeeva\unskip,\iMOSCOW}
\mbox{V. Eschenburg\unskip,\iMISSI}
\mbox{S. Fahey\unskip,\iCOLO}
\mbox{D. Falciai\unskip,\iFRAS}
\mbox{J.P. Fernandez\unskip,\iUCSC}
\mbox{K. Flood\unskip,\iMASS}
\mbox{R. Frey\unskip,\iOREG}
\mbox{E.L. Hart\unskip,\iTENN}
\mbox{K. Hasuko\unskip,\iTOHO}
\mbox{S.S. Hertzbach\unskip,\iMASS}
\mbox{M.E. Huffer\unskip,\iSLAC}
\mbox{X. Huynh\unskip,\iSLAC}
\mbox{M. Iwasaki\unskip,\iOREG}
\mbox{D.J. Jackson\unskip,\iRAL}
\mbox{P. Jacques\unskip,\iRUTG}
\mbox{J.A. Jaros\unskip,\iSLAC}
\mbox{Z.Y. Jiang\unskip,\iSLAC}
\mbox{A.S. Johnson\unskip,\iSLAC}
\mbox{J.R. Johnson\unskip,\iWISC}
\mbox{R. Kajikawa\unskip,\iNAGO}
\mbox{M. Kalelkar\unskip,\iRUTG}
\mbox{H.J. Kang\unskip,\iRUTG}
\mbox{R.R. Kofler\unskip,\iMASS}
\mbox{R.S. Kroeger\unskip,\iMISSI}
\mbox{M. Langston\unskip,\iOREG}
\mbox{D.W.G. Leith\unskip,\iSLAC}
\mbox{V. Lia\unskip,\iMIT}
\mbox{C. Lin\unskip,\iMASS}
\mbox{G. Mancinelli\unskip,\iRUTG}
\mbox{S. Manly\unskip,\iYALE}
\mbox{G. Mantovani\unskip,\iPERU}
\mbox{T.W. Markiewicz\unskip,\iSLAC}
\mbox{T. Maruyama\unskip,\iSLAC}
\mbox{A.K. McKemey\unskip,\iBRUN}
\mbox{R. Messner\unskip,\iSLAC}
\mbox{K.C. Moffeit\unskip,\iSLAC}
\mbox{T.B. Moore\unskip,\iYALE}
\mbox{M. Morii\unskip,\iSLAC}
\mbox{D. Muller\unskip,\iSLAC}
\mbox{V. Murzin\unskip,\iMOSCOW}
\mbox{S. Narita\unskip,\iTOHO}
\mbox{U. Nauenberg\unskip,\iCOLO}
\mbox{H. Neal\unskip,\iYALE}
\mbox{G. Nesom\unskip,\iOXF}
\mbox{N. Oishi\unskip,\iNAGO}
\mbox{D. Onoprienko\unskip,\iTENN}
\mbox{L.S. Osborne\unskip,\iMIT}
\mbox{R.S. Panvini\unskip,\iVAND}
\mbox{C.H. Park\unskip,\iSOONG}
\mbox{I. Peruzzi\unskip,\iFRAS}
\mbox{M. Piccolo\unskip,\iFRAS}
\mbox{L. Piemontese\unskip,\iFERR}
\mbox{R.J. Plano\unskip,\iRUTG}
\mbox{R. Prepost\unskip,\iWISC}
\mbox{C.Y. Prescott\unskip,\iSLAC}
\mbox{B.N. Ratcliff\unskip,\iSLAC}
\mbox{J. Reidy\unskip,\iMISSI}
\mbox{P.L. Reinertsen\unskip,\iUCSC}
\mbox{L.S. Rochester\unskip,\iSLAC}
\mbox{P.C. Rowson\unskip,\iSLAC}
\mbox{J.J. Russell\unskip,\iSLAC}
\mbox{O.H. Saxton\unskip,\iSLAC}
\mbox{T. Schalk\unskip,\iUCSC}
\mbox{B.A. Schumm\unskip,\iUCSC}
\mbox{J. Schwiening\unskip,\iSLAC}
\mbox{V.V. Serbo\unskip,\iSLAC}
\mbox{G. Shapiro\unskip,\iLBL}
\mbox{N.B. Sinev\unskip,\iOREG}
\mbox{J.A. Snyder\unskip,\iYALE}
\mbox{H. Staengle\unskip,\iCSU}
\mbox{A. Stahl\unskip,\iSLAC}
\mbox{P. Stamer\unskip,\iRUTG}
\mbox{H. Steiner\unskip,\iLBL}
\mbox{D. Su\unskip,\iSLAC}
\mbox{F. Suekane\unskip,\iTOHO}
\mbox{A. Sugiyama\unskip,\iNAGO}
\mbox{S. Suzuki\unskip,\iNAGO}
\mbox{M. Swartz\unskip,\iJHU}
\mbox{F.E. Taylor\unskip,\iMIT}
\mbox{J. Thom\unskip,\iSLAC}
\mbox{E. Torrence\unskip,\iMIT}
\mbox{T. Usher\unskip,\iSLAC}
\mbox{J. Va'vra\unskip,\iSLAC}
\mbox{R. Verdier\unskip,\iMIT}
\mbox{D.L. Wagner\unskip,\iCOLO}
\mbox{A.P. Waite\unskip,\iSLAC}
\mbox{S. Walston\unskip,\iOREG}
\mbox{A.W. Weidemann\unskip,\iTENN}
\mbox{E.R. Weiss\unskip,\iWASH}
\mbox{J.S. Whitaker\unskip,\iBU}
\mbox{S.H. Williams\unskip,\iSLAC}
\mbox{S. Willocq\unskip,\iMASS}
\mbox{R.J. Wilson\unskip,\iCSU}
\mbox{W.J. Wisniewski\unskip,\iSLAC}
\mbox{J.L. Wittlin\unskip,\iMASS}
\mbox{M. Woods\unskip,\iSLAC}
\mbox{T.R. Wright\unskip,\iWISC}
\mbox{R.K. Yamamoto\unskip,\iMIT}
\mbox{J. Yashima\unskip,\iTOHO}
\mbox{S.J. Yellin\unskip,\iUCSB}
\mbox{C.C. Young\unskip,\iSLAC}
\mbox{H. Yuta\unskip.\iAOMORI}

\it
  \vskip \baselineskip                   
  \centerline{(The SLD Collaboration)}   
  \vskip \baselineskip
  \baselineskip=.75\baselineskip   
\iAOMORI
  Aomori University, Aomori , 030 Japan, \break
\iBRI
  University of Bristol, Bristol, United Kingdom, \break
\iBRUN
  Brunel University, Uxbridge, Middlesex, UB8 3PH United Kingdom, \break
\iBU
  Boston University, Boston, Massachusetts 02215, \break
\iCOLO
  University of Colorado, Boulder, Colorado 80309, \break
\iCSU
  Colorado State University, Ft. Collins, Colorado 80523, \break
\iFERR
  INFN Sezione di Ferrara and Universita di Ferrara, I-44100 Ferrara, Italy, \break
\iFRAS
  INFN Lab. Nazionali di Frascati, I-00044 Frascati, Italy, \break
\iJHU
  Johns Hopkins University,  Baltimore, Maryland 21218-2686, \break
\iLBL
  Lawrence Berkeley Laboratory, University of California, Berkeley, California 94720, \break
\iMASS
  University of Massachusetts, Amherst, Massachusetts 01003, \break
\iMISSI
  University of Mississippi, University, Mississippi 38677, \break
\iMIT
  Massachusetts Institute of Technology, Cambridge, Massachusetts 02139, \break
\iMOSCOW
  Institute of Nuclear Physics, Moscow State University, 119899, Moscow Russia, \break
\iNAGO
  Nagoya University, Chikusa-ku, Nagoya, 464 Japan, \break
\iOREG
  University of Oregon, Eugene, Oregon 97403, \break
\iOXF
  Oxford University, Oxford, OX1 3RH, United Kingdom, \break
\iPERU
  INFN Sezione di Perugia and Universita di Perugia, I-06100 Perugia, Italy, \break
\iRAL
  Rutherford Appleton Laboratory, Chilton, Didcot, Oxon OX11 0QX United Kingdom, \break
\iRUTG
  Rutgers University, Piscataway, New Jersey 08855, \break
\iSLAC
  Stanford Linear Accelerator Center, Stanford University, Stanford, California 94309, \break
\iSOONG
  Soongsil University, Seoul, Korea 156-743, \break
\iTENN
  University of Tennessee, Knoxville, Tennessee 37996, \break
\iTOHO
  Tohoku University, Sendai 980, Japan, \break
\iUCSB
  University of California at Santa Barbara, Santa Barbara, California 93106, \break
\iUCSC
  University of California at Santa Cruz, Santa Cruz, California 95064, \break
\iVAND
  Vanderbilt University, Nashville,Tennessee 37235, \break
\iWASH
  University of Washington, Seattle, Washington 98105, \break
\iWISC
  University of Wisconsin, Madison,Wisconsin 53706, \break
\iYALE
  Yale University, New Haven, Connecticut 06511. \break

\rm
%

\end{center}
\begin{flushleft}
$^{*}$Deceased
\end{flushleft}


\begin{thebibliography}{99}
\bibitem{bclmix} See I. Dunietz, Fermilab-Pub-96/104-T. For the SLD sample, the
magnitude of this effect was estimated from the MC to increase the mean
mixing parameter for $b \rightarrow c (\bar c) \rightarrow l$ decays by 8.7\%.
\end{thebibliography}
\end{document}